\documentclass[sigconf]{acmart}

\AtBeginDocument{%
  }

\usepackage{tikz}
\usetikzlibrary{arrows.meta, positioning, shapes.geometric}

\usepackage[most]{tcolorbox}
\tcbset{
  colback=gray!5!white, 
  colframe=gray!50!black, 
  colbacktitle=black,
  coltitle=white,
  fonttitle=\bfseries\footnotesize,
  boxrule=0.5pt, 
  arc=3pt, 
  left=6pt, 
  right=6pt, 
  top=2pt, 
  bottom=2pt, 
  boxsep=4pt,
  enhanced,
}
\usepackage{cleveref}
\usepackage{paralist}
\usepackage{bm}
\usepackage[table]{xcolor}
\usepackage{colortbl}
\usepackage{stfloats}

\usepackage{booktabs}
\usepackage{multirow}
\usepackage{enumitem}

\copyrightyear{2025}
\acmYear{2025}
\setcopyright{acmlicensed}\acmConference[ICAIF '25]{6th ACM International Conference on AI in Finance}{November 15--18, 2025}{Singapore, Singapore}
\acmBooktitle{6th ACM International Conference on AI in Finance (ICAIF '25), November 15--18, 2025, Singapore, Singapore}
\acmDOI{10.1145/3768292.3770399}
\acmISBN{979-8-4007-2220-2/2025/11}

\begin{document}


\title{Aligning Language Models with Investor and Market Behavior for Financial Recommendations}
\author{Fernando Spadea}
\email{spadef@rpi.edu}
\orcid{0009-0006-4278-3666}
\affiliation{%
  \institution{Rensselaer Polytechnic Institute}
  \city{Troy}
  \state{New York}
  \country{USA}
}

\author{Oshani Seneviratne}
\email{senevo@rpi.edu}
\orcid{0000-0001-8518-917X}
\affiliation{%
  \institution{Rensselaer Polytechnic Institute}
  \city{Troy}
  \state{New York}
  \country{USA}
}


\begin{abstract}
Most financial recommendation systems often fail to account for key behavioral and regulatory factors, leading to advice that is misaligned with user preferences, difficult to interpret, or unlikely to be followed. We present FLARKO (Financial Language-model for Asset Recommendation with Knowledge-graph Optimization), a novel framework that integrates Large Language Models (LLMs), Knowledge Graphs (KGs), and Kahneman-Tversky Optimization (KTO) to generate asset recommendations that are both profitable and behaviorally aligned. FLARKO encodes users’ transaction histories and asset trends as structured KGs, providing interpretable and controllable context for the LLM.
To demonstrate the adaptability of our approach, we develop and evaluate both a centralized architecture (CenFLARKO) and a federated variant (FedFLARKO). 
To our knowledge, this is the first demonstration of combining KTO for fine-tuning of LLMs for financial asset recommendation. We also present the first use of structured KGs to ground LLM reasoning over behavioral financial data in a federated learning (FL) setting.
Evaluated on the FAR-Trans dataset, FLARKO consistently outperforms state-of-the-art recommendation baselines on behavioral alignment and joint profitability, while remaining interpretable and resource-efficient.
\end{abstract}



\maketitle

\section{Introduction}


Many existing financial asset recommendation systems, while valuable, often fall short in real-world settings because financial decision-making is influenced by more than just numerical optimization. As highlighted by \citet{sanz2024far} and \citet{lee2024stock}, individuals frequently disregard theoretically optimal financial advice if it conflicts with their personal preferences, ethical views, or logistical constraints. Thus, aligning recommendations with user preferences is fundamental to the recommendation system's effectiveness. Simply maximizing expected profitability is insufficient if the recommendations are unlikely to be followed.

Conventional recommendation models are further limited by their rigid architectures and reliance on static user profiles or historical return patterns. This makes them poorly suited to capture the nuanced, evolving nature of individual investor behavior, including preferences that shift over time or depend on non-financial factors. 
Such constraints are particularly pronounced in regulated financial environments, where data centralization may not be feasible due to legal, geographic, or institutional boundaries. In these cases, a federated learning strategy offers a promising alternative by enabling collaborative model training across institutions while preserving data locality. Since these federated scenarios inherently involve client data that is not independent and identically distributed (non-IID), measuring the effects of this data distribution is also important.
%
Recent advances in large language models (LLMs) have unlocked new possibilities for personalized decision support, yet their adoption in regulated financial domains remains limited. A key barrier is the tension between personalization and compliance: most financial institutions cannot centralize sensitive client data due to consumer privacy regulations, and vanilla LLMs offer little transparency or behavioral grounding.

In this paper, we propose a novel system that uses an LLM to generate asset recommendations based on both \textbf{user behavioral data} that reflects user's preferences, patterns, and intent (e.g., customer transaction history) and \textbf{non-behavioral market signals} (e.g., asset price history). 
We structure both types of these data as knowledge graphs (KGs), which are passed into the LLM context to enable more informed and interpretable recommendations; KGs provide a structured, interconnected, and semantically rich representation of financial behavior and market trends, allowing the model to reason over personalized, contextual, and external signals. 
%

For aligning LLM recommendations with user preferences, we leverage Kahneman-Tversky Optimization (KTO)~\cite{ethayarajh2024kto}. 
We select KTO for its computational efficiency, behavioral grounding, and suitability for distributed optimization. In particular, KTO performs well in federated learning environments~\cite{spadea2025federated}, where user data is siloed and sensitive.
KTO requires only a binary desirability label (indicating whether a recommendation was both profitable and consistent with user behavior), making it significantly easier to collect alignment data compared to ranking or pairwise preference approaches. 

\subsection{Contributions}

\begin{figure*}[t]
    \centering
    \includegraphics[width=1\linewidth]{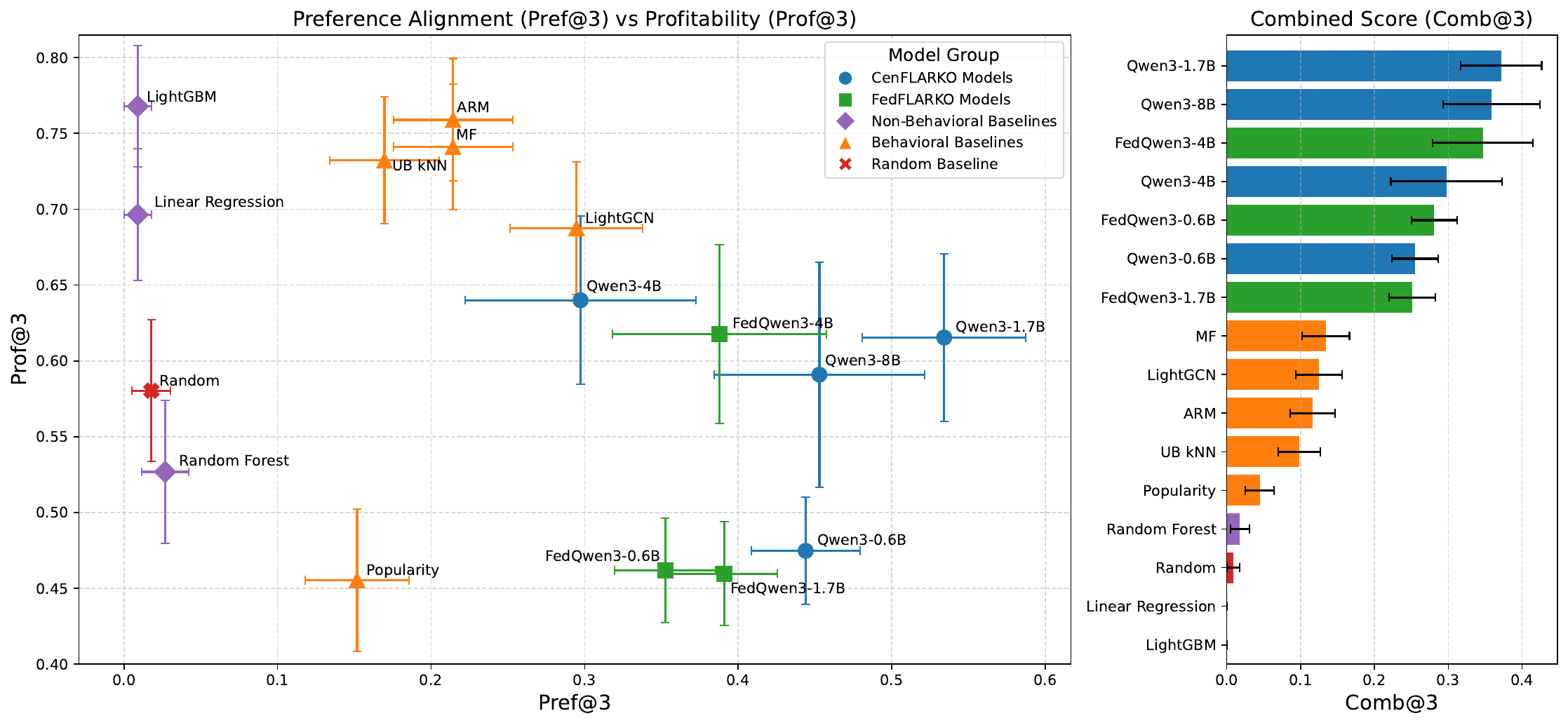}
    \caption{Performance Comparison of \textbf{CenFLARKO} and \textbf{FedFLARKO} Against Baseline Models \\
\normalfont\small 
The left panel plots preference alignment (Pref@3) against profitability (Prof@3) for all models. Prof@3 captures the ability to recommend assets that generate positive returns over the next 180 days, while Pref@3 reflects behavioral alignment by checking whether the user actually purchased the recommended assets. Although some models (e.g., LightGBM) achieve higher profitability, FLARKO-based models, particularly \textbf{CenFLARKO}, excel in preference alignment, demonstrating stronger user-centric performance. The right panel reports Comb@3, which quantifies how often the model recommends assets that are both profitable and behaviorally aligned, representing actionable, high-quality financial advice. Both \textbf{CenFLARKO} and \textbf{FedFLARKO} outperform all baselines on this crucial metric, validating the strength of our approach in real-world financial asset recommendation scenarios.}

    \label{fig:baselines}
\end{figure*}

In this work, we present FLARKO (Financial Language-model for Asset Recommendation with Knowledge-graph Optimization), a unified framework that combines LLMs with structured KGs to deliver personalized, behaviorally aligned financial asset recommendations. We apply this core framework in both centralized (CenFLARKO) and federated (FedFLARKO) settings, demonstrating its versatility across deployment environments with varying privacy and data-sharing constraints.

Our methodology is tested using the FAR-Trans dataset~\cite{sanz2024far}, and as illustrated in \Cref{fig:baselines}, our experiments demonstrate competitive performance and key advantages over the state-of-the-art recommendation system baselines.
Our specific contributions include:


\begin{enumerate}
    \item \textbf{A unified LLM-KG framework:} FLARKO integrates LLMs with personalized behavioral and market-level KGs, enabling contextual reasoning about financial assets. 
    The KGs provide structured, interpretable representations of user behavior and market dynamics, while the LLM leverages this symbolic context to generate flexible, natural language asset recommendations. This combination allows FLARKO to support user-centric constraints, ethical investment rules, and portfolio diversification strategies~\cite{spadea2025avoiding, spadea2025bursting}.

    \item \textbf{User preference alignment in financial asset recommendations:} We are the first to show that aligning LLM-generated financial recommendations with actual user investment behavior, measured via Pref@3 and Comb@3, can be effectively achieved in both centralized and federated settings.
    As shown in \Cref{fig:baselines}, FLARKO significantly outperforms many recommendation baselines, validating FLARKO's core design principle that actionable recommendations must be both profitable and behaviorally aligned.


    \item \textbf{Efficient and cost-effective performance with mid-sized LLMs:} We evaluate our framework across a range of LLMs (0.6B to 8B parameters) and demonstrate that state-of-the-art performance is attainable without relying on massive, resource-intensive models. Our empirical results show that performance does not strictly increase with model size; in fact, models in the 1.7B to 4B parameter range often delivered the best results, even outperforming the larger 8B model. 
    This illustrates that FLARKO offers a practical and resource-efficient solution for real-world deployment in financial applications.



    \item \textbf{Federated collaboration under realistic data constraints:} We introduce FedFLARKO, a framework for collaboratively training financial asset recommendation models across multiple institutions without sharing sensitive or proprietary data, addressing key regulatory and competitive barriers. Through extensive evaluation under both IID and non-IID clients, we show that FedFLARKO remains robust, and even improves in performance with larger models, under realistic, heterogeneous client scenarios, making it well-suited for real-world federated deployments in finance.
    
\end{enumerate}

 \subsection{Use Cases}


FLARKO is designed to operate in both centralized and federated environments, making it applicable across a wide range of financial use cases. 

In centralized deployments, such as within a single financial institution, FLARKO can serve as a tool for personalized wealth management. For example, a private bank or advisory firm can use CenFLARKO to generate behaviorally aligned investment recommendations that incorporate transaction history, risk preferences, and the firm's ethical constraints. Advisors can interact with and override these recommendations.

In federated settings, FLARKO enables collaboration across institutions without data centralization. For example, a consortium of banks or financial platforms operating in different jurisdictions can use FedFLARKO to jointly improve their recommendation systems without sharing sensitive customer data. This supports compliance with data privacy regulations like General Data Protection Regulation (GDPR) or California Consumer Privacy Act (CCPA) while enabling cross-institutional learning.

\section{Related Work}


\paragraph{\textbf{Financial Asset Recommendation Systems:}}
These have historically relied on quantitative models and rule-based expert systems~\cite{markowitz2010portfolio,gum1987expert}. These traditional approaches, along with collaborative filtering (which predicts user preferences by analyzing similarities between users or items), content-based filtering (which recommends items similar to those a user has liked in the past), and hybrid systems (which combine multiple approaches), have been foundational in broader recommender systems (e.g., for e-commerce and information retrieval). However, their direct application to the high-stakes financial sector reveals a set of inherent limitations.
These challenges are often exacerbated by the unique characteristics of financial data, the specific nature of user behavior in financial contexts, and the dynamic shifts within financial markets~\cite{zibriczky2016recommender}. Consequently, these traditional systems inherently struggle to capture nuanced, non-numerical financial relationships and adapt to dynamic market shifts, limiting their effectiveness for sophisticated financial recommendations. Our work, in contrast, is specifically designed to overcome these shortcomings by integrating LLMs with dynamic user behavior-oriented PKGs to process complex user-specific financial information and also respond to real-time market changes.

\paragraph{\textbf{LLMs in Finance:}}
LLMs have emerged as powerful tools in finance due to their ability to understand complex financial text (such as news articles, earnings reports, and social media sentiments), capture nuanced sentiment, and perform reasoning over both structured and unstructured data. 
Financial LLMs, including FinBERT~\cite{liu2021finbert}, BloombergGPT~\cite{wu2023bloomberggpt}, FinGPT~\cite{liu2023fingpt}, InvestLM~\cite{yang2023investlm}, and FinLlama~\cite{iacovides2024finllama}, have demonstrated applications in areas such as sentiment analysis, market forecasting, and risk assessment. However, these are often monolithic models, and there are inherent challenges of using LLMs in high-stakes private financial contexts. Additionally, the tendency of LLMs to generate plausible but incorrect information (hallucination) could have severe consequences in financial recommendations. Furthermore, LLMs are predominantly trained on historical data, necessitating a robust methodology to handle rapidly changing financial markets and real-time information. 
LLMs can also inherit biases from their training data, potentially leading to unfair or discriminatory financial advice.
We address all these issues via the incorporation of KGs to guide the LLM in a context-aware manner. 

\paragraph{\textbf{KGs in Financial AI:}}

The open-source Financial Dynamic Knowledge Graph (FinDKG)~\cite{li2024findkg} models global economic and market trends, with applications in risk management, thematic investing, and economic forecasting. However, while FinDKG focuses on macroeconomic trends and general financial intelligence, our work distinctively integrates LLMs with KGs for personalized financial asset recommendation, emphasizing behavioral alignment and optimizing for individual user compliance alongside profitability.

\paragraph{\textbf{Behavioral Aspects in Financial Asset Recommendations:}}
\citet{sanz2024far} introduced the FAR-Trans dataset, which captures anonymized customer transaction histories alongside asset price data. Their evaluation focuses on two key metrics: (1) the profitability of recommended assets over a 6-month horizon and (2) alignment between recommendations and actual user behavior. This dual evaluation reflects a critical insight: financial advice is more likely to be followed when it aligns with the user's preferences and past actions.
\citet{lee2024stock} reinforce this principle, demonstrating that behaviorally aligned recommendations using the FAR-Trans dataset, even if suboptimal in terms of pure returns, can lead to higher investor adoption than purely profit-maximizing strategies. 
We build directly on these insights, aiming to improve both profitability and behavioral alignment through a unified framework that integrates LLMs with structured KG inputs.


\section{FLARKO Data Architecture}

A core component of the FLARKO data architecture is its use of KGs to encode financial context in a form that LLMs can interpret and reason over.



\subsection{KG Design}

LLMs are powerful tools for understanding and generating natural language, but when applied to structured decision-making tasks, they require explicit contextual grounding to ensure interpretability, consistency, and robustness~\cite{lewis2020retrieval}. In FLARKO, we address this need by encoding user transaction histories and asset price information into structured KGs, which serve as symbolic inputs to the LLM during inference. This grounding enables the model to reason over both personalized behavioral signals and market-level financial trends in a transparent and controllable manner, while also mitigating common LLM pitfalls such as hallucination by anchoring generation to factual, structured inputs. To ensure scalability, our KG construction strategy balances expressiveness with efficiency, capturing essential financial context without exceeding the LLM’s token limitations.

Each recommendation instance in FLARKO is grounded in two distinct KGs:  
\begin{enumerate}[leftmargin=1.5em]
    \item A \textbf{Personal Knowledge Graph (PKG):} Encodes an investor's past transaction behavior, capturing asset interactions over time and serving as a proxy for user intent and preferences.
    \item A \textbf{Market Knowledge Graph (MKG):} Encodes external financial signals, including asset price trends, and sector metadata, derived from historical price series and asset descriptors.
\end{enumerate}

Both KGs are constructed using the standard subject-predicate-object triple format, a widely adopted formalism for structured knowledge representation~\cite{hogan2021knowledge}. 
To interface effectively with the LLM, we serialize these KGs into \texttt{JSON-LD}~\cite{sporny2014json}, a format that balances machine interpretability with LLM-friendliness due to its semantic structure and compatibility with web standards. Using JSON-LD also allows for easy integration of financial domain ontologies.

To meet the token budget constraints of large language models, each recommendation instance—or \emph{data point}—is built with a cutoff timestamp, denoted as \texttt{RECOMMENDATION\_DATE}. This defines the historical window from which the user’s transactions and relevant market summaries are extracted. To reduce input length while preserving semantic richness, we compress the raw data by aggregating time-series signals (e.g., rolling price statistics over 10-week intervals) and pruning redundant triples. The resulting KG pair is capped at 5,000 triples to ensure it fits within the LLM’s context window during prompt construction.

\paragraph{\textbf{PKG Construction}}

To build the PKG for a given user, we extract essential features from their transaction history, including the  International Securities Identification Number (ISIN) of the assets held, transaction type (buy/sell), transaction value, and timestamp.
Redundant or non-informative fields are omitted to minimize token overhead while retaining the most semantically meaningful information. These transaction records are filtered to include only those that occurred prior to the \texttt{RECOMMENDATION\_DATE}, which serves as the cutoff point for historical context.

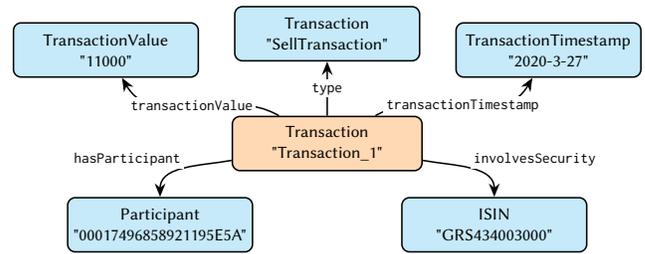
\begin{figure}[h]
    \centering
    \resizebox{\columnwidth}{!}{%


\begin{tikzpicture}[
    node distance=0.7cm and 0.6cm,
    entity/.style={
        rectangle, rounded corners, draw, thick, fill=orange!30,
        minimum width=3.5cm, minimum height=1cm, align=center, font=\sffamily
    },
    attribute/.style={
        rectangle, rounded corners, draw, thick, fill=cyan!20,
        text width=3.2cm, minimum height=1cm, align=center, font=\sffamily
    },
    edge_label/.style={
        fill=white, inner sep=1.5pt, font=\small\ttfamily
    },
    arrow/.style={
        -{Stealth[length=2.5mm, width=2mm]}, thick
    }
]

\node[entity] (Transaction) {Transaction\\"Transaction\_1"};

\node[attribute, left=of Transaction, xshift=1cm, yshift=-1.5cm] (hasParticipant) {Participant\\"00017496858921195E5A"};
\node[attribute, right=of Transaction, xshift=-1cm, yshift=-1.5cm] (involvesSecurity) {ISIN\\"GRS434003000"};
\node[attribute, above left=of Transaction] (TransactionValue) {TransactionValue\\"11000"};
\node[attribute, above right=of Transaction] (TransactionTimestamp) {TransactionTimestamp\\"2020-3-27"};
\node[attribute, above=1cm of Transaction] (type) 
{Transaction \\"SellTransaction"};

\path[arrow] (Transaction) edge[out=190, in=90] node[edge_label, auto, swap] {hasParticipant} (hasParticipant);
\path[arrow] (Transaction) edge[out=-10, in=90] node[edge_label, auto] {involvesSecurity} (involvesSecurity);
\path[arrow] (Transaction) edge[out=150, in=-60] node[edge_label, pos=0.5] {transactionValue} (TransactionValue);
\path[arrow] (Transaction) edge[out=30, in=-120] node[edge_label, pos=0.5] {transactionTimestamp} (TransactionTimestamp);
\path[arrow] (Transaction) edge node[edge_label] {type} (type);

\end{tikzpicture}

    }
    \caption{Example user transaction in the PKG. 
    }
    \label{fig:PKG-schema}
\end{figure}

\Cref{fig:PKG-schema} illustrates an exemplar transaction entry associated with a user (i.e., a \texttt{Participant})  within a PKG. The central entity, shown in orange, represents a specific transaction instance (e.g., \texttt{"Transaction\_1"}). This transaction node is linked to a set of typed attribute nodes (shown in blue), each capturing a distinct feature: the transaction type (e.g., \texttt{"SellTransaction"}), the monetary value of the trade (e.g., \texttt{"11000"}), the timestamp (e.g., \texttt{"2020-3-27"}), the financial instrument involved (identified via ISIN), and the participant associated with the transaction. These attributes are connected to the central transaction entity via well-defined predicates (e.g., \texttt{transactionValue}, \texttt{involvesSecurity}, \texttt{hasParticipant}).

\paragraph{\textbf{MKG Construction}}


To represent broader market-level financial signals, FLARKO constructs an MKG that encodes historical asset performance and descriptive metadata. Rather than including every raw price point, we aggregate price data into \textit{TenWeekPriceSummary} entities, each summarizing an asset’s behavior over a fixed 10-week interval, where only the summary periods ending before the selected \texttt{RECOMMENDATION\_DATE} are included.
While the ten-week aggregation period serves as a default summarization cadence, FLARKO can be designed to flexibly incorporate alternative temporal resolutions to adapt to varying user profiles and recommendation contexts, where finer or coarser market summaries may be more appropriate for aligning with user intent or portfolio strategy.

\begin{figure}[h]
    \centering
    \resizebox{\columnwidth}{!}{%



\begin{tikzpicture}[
    node distance=0.7cm and 0.8cm,
    entity/.style={
        rectangle, rounded corners, draw, thick, fill=orange!30,
        minimum width=3.5cm, minimum height=1cm, align=center, font=\sffamily
    },
    attribute/.style={
        rectangle, rounded corners, draw, thick, fill=cyan!20,
        text width=2.8cm, minimum height=1cm, align=center, font=\sffamily
    },
    edge_label/.style={
        fill=white, inner sep=1.5pt, font=\small\ttfamily
    },
    arrow/.style={
        -{Stealth[length=2.5mm, width=2mm]}, thick
    }
]

\node[entity] (summary) {TenWeekPriceSummary\\"TenWeekPriceSummary\_1"};

\node[attribute, left=of summary] (PeriodEndPrice) {PeriodEndPrice\\"8.54"};
\node[attribute, right=of summary] (PeriodAvgPrice) {PeriodAveragePrice\\"9.1679792"};
\node[attribute, above=of PeriodEndPrice] (PeriodHighPrice) {PeriodHighPrice\\"9.5"};
\node[attribute, above=of PeriodAvgPrice] (PeriodLowPrice) {PeriodLowPrice\\"8.54"};
\node[attribute, above=1cm of summary] (PeriodEndDate) {PeriodEndDate\\"2018-05-27"};

\node[entity, below=1.5cm of summary] (asset) {Asset\\"Asset\_1"};

\node[attribute, left=of asset] (identifier) {ISIN\\"GRS495003006"};
\node[attribute, right=of asset] (category) {Asset Category\\"Stock"};
\node[attribute, below=of identifier, xshift=1.5cm] (sector) {Sector\\"Industrials"};
\node[attribute, below=of category, xshift=-1.5cm] (industry) {Industry\\"Airlines"};

\path[arrow] (summary) edge node[edge_label, swap] {priceOf} (asset);

\path[arrow] (summary) edge[out=-160, in=-90] node[edge_label] {periodEndPrice} (PeriodEndPrice);
\path[arrow] (summary) edge[out=-20, in=-90] node[edge_label, swap] {periodAveragePrice} (PeriodAvgPrice);
\path[arrow] (summary) edge[out=140, in=-60] node[edge_label, pos=0.5] {periodHighPrice} (PeriodHighPrice);
\path[arrow] (summary) edge[out=40, in=-120] node[edge_label, swap, pos=0.5] {periodLowPrice} (PeriodLowPrice);
\path[arrow] (summary) edge node[edge_label] {periodEndDate} (PeriodEndDate);

\path[arrow] (asset) edge[out=170, in=90] node[edge_label, pos=0.4] {identifier} (identifier);
\path[arrow] (asset) edge[out=10, in=90] node[edge_label, pos=0.4] {category} (category);
\path[arrow] (asset) edge[out=-120, in=90] node[edge_label] {sector} (sector);
\path[arrow] (asset) edge[out=-60, in=90] node[edge_label, swap] {industry} (industry);

\end{tikzpicture}

    }
    \caption{An example asset summary in the MKG.}
    \label{fig:MKG-schema}
\end{figure}
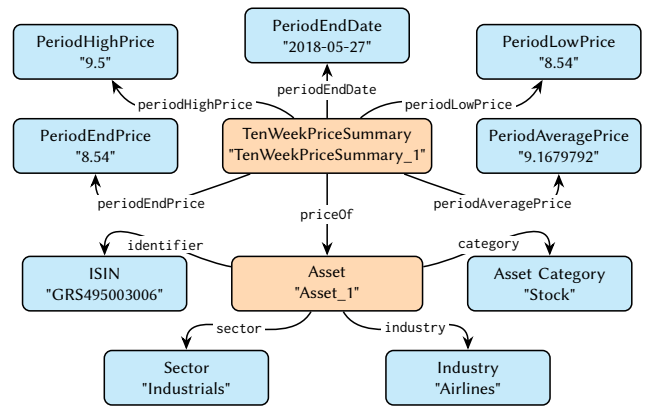

As shown in \Cref{fig:MKG-schema}, each price summary node (in orange) includes attributes such as the period's high, low, average, and end prices, each represented as a typed literal node (in blue).
Each \textit{TenWeekPriceSummary} is linked to an associated asset node, which is itself enriched with relevant metadata: ISIN, asset category, sector, and industry.

\subsection{LLM Prompt Design}

We employ FLARKO with LLMs from the Qwen family~\cite{huggingface2024qwen3} and provide them with carefully designed prompts that incorporate PKG and MKG content. 
The prompts are structured with placeholders where \texttt{\{...\}} represents dynamic inputs to the LLM, and \texttt{[...]} denotes placeholders that the model is expected to complete. Each recommendation interaction begins with a system prompt that establishes the LLM’s role in the recommendation task:

\begin{tcolorbox}[title=System Prompt, coltitle=white, fonttitle=\bfseries, sharp corners=southwest, enhanced]
\ttfamily\small
You are an expert financial analyst AI. Your task is to analyze a user's transaction history and supplementary market data to provide personalized asset recommendations. The user will ask for recommendations for the next 180 days from a given "current date".

You MUST provide your response in the following format, and only this format:

[An introductory sentence]\\
- [ASSET\_ISIN\_1]\\
- [ASSET\_ISIN\_2]\\
- [ASSET\_ISIN\_3]
\end{tcolorbox}

Then, the asset price history is provided as follows:

\begin{tcolorbox}[title=Asset Price History Information, coltitle=white, fonttitle=\bfseries, sharp corners=southwest, enhanced]
\ttfamily\small
Here is the supplementary knowledge graph with asset information and historical prices in JSON-LD format:

```jsonld\\
\{MARKET\_KNOWLEDGE\_GRAPH\}\\
```
\end{tcolorbox}

Then, the user's transaction history is provided:

\begin{tcolorbox}[title=User's Transaction History Information, coltitle=white, fonttitle=\bfseries, sharp corners=southwest, enhanced]
\ttfamily\small
Here is the user's transaction history in JSON-LD format:

```jsonld\\
\{PERSONAL\_KNOWLEDGE\_GRAPH\}\\
```
\end{tcolorbox}

Finally, the user's request would be presented as a user message.

\begin{tcolorbox}[title=User Request Template, coltitle=white, fonttitle=\bfseries, sharp corners=southwest, enhanced]
\ttfamily\small
Considering all the provided data, and assuming the current date is \{RECOMMENDATION\_DATE\}, please provide a list of asset recommendations for my portfolio for the next 6 months.
\end{tcolorbox}

\section{Behavioral Alignment}

To ensure that FLARKO's recommendations align with actual user preferences and investment behavior, we fine-tune the LLM using KTO~\cite{ethayarajh2024kto}, a lightweight, behaviorally motivated alignment method. KTO requires only binary feedback on whether a model's recommendation is desirable or not, making it particularly suitable for federated learning settings, where granular supervision is difficult to obtain. 

\paragraph{\textbf{KTO Data Design}}


KTO's key advantage lies in its minimal supervision requirements. Each training data point consists of:

\begin{enumerate}[topsep=2pt, itemsep=2pt, parsep=0pt]
\item \textbf{Prompt}: Natural language input or a truncated user–LLM conversation.
\item \textbf{Completion}: A potential response to the prompt.
\item \textbf{Label}: A binary signal indicating whether the completion is desired or not.
\end{enumerate}

This simplicity makes KTO highly scalable and cost-effective, particularly in federated settings where users span diverse profiles and detailed annotations are impractical~\cite{spadea2025federated}. Notably, prior work has shown KTO to outperform more complex preference optimization methods~\cite{saeidi2024insights}.
\begin{sloppypar}
To construct these training data points, we first select a \texttt{RECOMMENDATION\_DATE}, then gather the user’s transaction history (PKG) and asset summaries (MKG) leading up to that date to build the KG context. Using this context, we generate one or more candidate recommendations and label them based on whether they match the user's actual behavior and market outcomes.
\end{sloppypar}

\paragraph{\textbf{Labeling Recommendations for Alignment Training}}

To support alignment training with KTO, we label recommendations as either \textit{desirable} or \textit{undesirable} based on two key criteria: behavioral alignment and financial performance. An asset is labeled as \textbf{desirable} if it satisfies both of the following: (i) the user actually purchased the asset within the 180-day window following the \texttt{RECOMMENDATION\_DATE}, and (ii) the asset delivered a positive return over the same period. This intersection ensures that selected recommendations are not only aligned with the user’s preferences but also financially sound. Conversely, assets that were either not purchased or resulted in negative returns are labeled as \textbf{undesirable}, and serve as negative examples during alignment training.

For each prompt, we generate two data points: one containing a completion with desirable recommendations and the other with undesirable ones, based on the labeling criteria described above. Each completion includes up to 20 asset recommendations.

\section{Federated Learning Setup}

To enable collaborative training and fine-tuning of the LLM across institutions, FLARKO supports a decentralized (FedFLARKO) training mode. FedFLARKO, allows clients to train locally on their own PKGs, sharing only model updates with a central aggregator.

\paragraph{\textbf{Federated Client Modeling}}
\label{subsec:client}

To simulate realistic federated deployments, each client in FedFLARKO represents a financial institution or branch serving a unique customer base. Our client design accounts for heterogeneity in investor profiles by assigning each client a distinct behavioral distribution over user types, risk preferences, and investment capacities. This modeling reflects real-world scenarios where different firms serve demographically or behaviorally distinct investor populations, resulting in non-IID local data.
To evaluate generalization under diverse data conditions, we simulate both skewed (non-IID) and randomly distributed (IID) client assignments.


\paragraph{\textbf{Communication-Efficient Coordination}}


FedFLARKO employs communication-efficient federated learning via low rank adaptation of LLMs (LoRA)-based parameter-efficient tuning~\cite{hu2022lora}. Additionally, 4-bit quantization is implemented to significantly decrease parameter size.
The use of LoRA and quantization reduces communication costs while preserving strong alignment performance, as quantified in \Cref{sec:evaluation-comm}.

\section{Evaluation}
\label{sec:evaluation}

To evaluate FLARKO's performance in both centralized and federated contexts, we conduct a comprehensive experimental study using the FAR-Trans dataset~\cite{sanz2024far}. Because the FAR-Trans dataset consists of several markets, our experimental results here should generalize to the other datasets and markets. This section describes the test data, experimental protocol, client simulation, baseline comparisons, and metrics used.

\subsection{Dataset and Experimental Timeline}
\label{subsec:data-splits}

FAR-Trans dataset~\cite{sanz2024far} contains customer transaction histories, asset price histories, and investor profile information.

The training dataset is from January 2, 2018, to November 30, 2021, and includes prompts sampled every four weeks from August 1, 2019, to June 1, 2021 (180 days prior to November 30, 2021). 
Then, the test dataset is between December 1, 2021, and November 29, 2022, with test prompts being constructed every two weeks from December 1, 2021, to June 2, 2022 (180 days prior to November 29, 2022). We use the 180-day buffers to determine the desirable recommendations.



Each test instance includes:
\begin{itemize}
    \item A user prompt with historical PKG and MKG inputs.
    \item A list of \textit{purchased assets} in the 180 days following the\\ \texttt{RECOMMENDATION\_DATE}.
    \item A list of \textit{profitable assets} over the same period.
    \item A list of \textit{desirable assets} (intersection of the above).
\end{itemize}

\subsection{Baseline Comparisons}
\label{subsec:baselines}

We compare FLARKO against baselines from the FAR-Trans benchmark:
\begin{itemize}
    \item \textbf{Asset price-based:} Random Forest, Linear Regression, LightGBM~\cite{ke2017lightgbm}
    \item \textbf{Behavior-based:} Popularity, LightGCN~\cite{he2020lightgcn}, ARM~\cite{agrawal1994r}, MF~\cite{rendle2020neural}, UB kNN~\cite{nikolakopoulos2021trust}
    \item \textbf{Random:} Uniform random sampling
\end{itemize}


\subsection{Metrics}
\label{subsec:metrics}

Performance is measured using Hits@3, with three evaluation variants:
\begin{itemize}
    \item \textbf{Pref@3:} Hit rate against \textit{purchased assets} 
    \item \textbf{Prof@3:} Hit rate against \textit{profitable assets} 
    \item \textbf{Comb@3:} Hit rate against assets that are both purchased and profitable (\textit{desirable assets}
\end{itemize}

\subsection{Federated Client Simulation}
\label{subsec:client-simulation}

We simulate 20 clients representing financial institutions. Clients are defined based on three user-level attributes available in the FAR-Trans dataset~\cite{sanz2024far}:
\begin{itemize}
    \item \textbf{Customer Type:} Mass, Premium, Legal Entity, Professional
    \item \textbf{Risk Level:} Conservative, Moderate, Aggressive
    \item \textbf{Investment Capacity:} e.g., \$30K, \$80K
\end{itemize}


To model heterogeneity, each client is assigned a synthetic non-IID profile using randomized probability distributions over the above attributes. This results in realistic behavioral skew (e.g., some clients serving mostly conservative, low-capacity customers). As a control, we also run experiments in an IID setting where clients receive uniform distributions.

Across all clients, we generate 23,784 labeled prompt-completion examples for KTO fine-tuning, evenly split between desirable and undesirable completions.

\subsection{Model Configurations and Training Protocol}
\label{subsec:training-setup}

We use various sizes of the Qwen3 model (0.6B, 1.7B, 4B, and 8B)~\cite{huggingface2024qwen3}. 
The 8B model is excluded from FL due to compute constraints.
These models usually have a context length of 32,768 tokens, but we use yarn~\cite{peng2023yarn} to increase it to 131,072. 

For fine-tuning, we use LoRA~\cite{hu2022lora} with a rank of 16 and alpha of 64. The rank controls the size of the LoRA adaptations, and we choose a rank of 16 because it provides enough new parameters to work with while keeping the fine-tuning lightweight. The alpha controls the influence of the LoRA adaptations, and we choose a high value to make sure the model learns the desired behaviors, 
but we do not choose an even higher value to avoid nullifying the properties of the base model. This makes fine-tuning more efficient as fewer parameters need to be trained. Additionally, we use 4-bit quantization to minimize their VRAM usage. 

For centralized training, we run for 3 epochs. 
In the federated setting, to manage communication overhead and client computational load, in each of 200 communication rounds, 3 clients are randomly selected from the 20 available in the pool to perform local updates for 0.1 epochs. This configuration ensures that, on average, each client participates in approximately 30 rounds, resulting in a total training equivalent to about 3 epochs per client, thereby aligning the overall training effort with the centralized learning baseline.

\subsection{Communication Overhead in FL}
\label{sec:evaluation-comm}


FLARKO’s federated training leverages LoRA-based parameter-efficient fine-tuning, which substantially reduces communication costs. This is because only the LoRA adapter weights (rather than full model parameters) are exchanged. In each round:

\begin{itemize}
    \item A random subset of 3 out of the 20 simulated clients is selected for training.
    \item Each selected client uploads its local LoRA adapter weights to the central aggregator.
    \item The server aggregates the updates and broadcasts a new global model to all clients.
\end{itemize}

Thus, the server's total communication cost per round is 23 times the LoRA adapter size (3 client downloads + 20 client uploads), while each client incurs a single download and the selected clients incur an additional upload. \Cref{tab:lora_adapter_sizes} details the corresponding LoRA adapter sizes of the models we tested. Even for the largest model tested (Qwen3-8B), the per-round communication overhead only reaches 478.69 MB.

\begin{table}[h!]
\centering
\caption{LoRA Adapter Sizes for Qwen3 Models}
\label{tab:lora_adapter_sizes}
\begin{tabular}{lrr}
\toprule
\textbf{Model} & \textbf{Trainable Parameters} & \textbf{Adapter Size (4-bit)} \\
\midrule
Qwen3-0.6B     & 10,092,544                    & 4.8125 MB               \\
Qwen3-1.7B     & 17,432,576                    & 8.3125 MB               \\
Qwen3-4B       & 33,030,144                    & 15.75 MB                \\
Qwen3-8B       & 43,646,976                    & 20.8125 MB              \\
\bottomrule
\end{tabular}
\end{table}

\section{Results}

\subsection{CenFLARKO Results}

\begin{table}[t]
\centering
\caption{
Performance of CenFLARKO across different model sizes and input configurations\\
\normalfont\small Results are presented as mean $\pm$ standard error of a proportion. The best results for each model are in \textbf{bold}, and the best overall are marked with a $\uparrow$.
}
\label{tab:supp}
\footnotesize 
\setlength{\tabcolsep}{3pt}
\begin{tabular}{l@{\hskip 4pt}l@{\hskip 1pt}c@{\hskip 1pt}c@{\hskip 1pt}c}
\toprule
\textbf{Model} & \textbf{Data} & \textbf{Pref@3} & \textbf{Prof@3} & \textbf{Comb@3} \\
\midrule

\multirow{4}{*}{Qwen3-0.6B}
    & Combined        & $0.0354 \pm 0.0131$          & $0.1010 \pm 0.0214$          & $0.0152 \pm 0.0087$ \\
    & PKG    & $\bm{0.4439 \pm 0.0355}$ & $0.4694 \pm 0.0356$          & $\bm{0.2551 \pm 0.0311}$ \\
    & MKG    & $0.4141 \pm 0.0350$          & $\bm{0.4747 \pm 0.0355}$ & $0.2323 \pm 0.0300$ \\
    & Nothing         & $0.4352 \pm 0.0357$          & $0.4219 \pm 0.0356$          & $0.2240 \pm 0.0301$ \\
    
\midrule
\multirow{4}{*}{Qwen3-1.7B}
    & Combined        & $0.0990 \pm 0.0216$          & $0.4975 \pm 0.0354$          & $0.0524 \pm 0.0161$ \\
    & PKG    & $0.4434 \pm 0.0483$          & $0.4528 \pm 0.0483$          & $0.2642 \pm 0.0428$ \\
    & MKG    & $\bm{0.5341 \pm 0.0532}$$\uparrow$ & $0.5169 \pm 0.0530$ & $0.3448 \pm 0.0510$ \\
    & Nothing         & $0.5000 \pm 0.0566$          & $\bm{0.6154 \pm 0.0551}$ & $\bm{0.3718 \pm 0.0547}$$\uparrow$ \\

\midrule
\multirow{4}{*}{Qwen3-4B}
    & Combined        & $0.2740 \pm 0.0522$          & $\bm{0.6400 \pm 0.0554}$$\uparrow$ & $0.1644 \pm 0.0434$ \\
    & PKG    & $\bm{0.2973 \pm 0.0751}$ & $0.4324 \pm 0.0814$          & $\bm{0.2973 \pm 0.0751}$ \\
    & MKG    & $0.1250 \pm 0.0523$          & $0.1500 \pm 0.0565$          & $0.0750 \pm 0.0416$ \\
    & Nothing         & $0.1795 \pm 0.0615$          & $0.1316 \pm 0.0548$          & $0.1316 \pm 0.0548$ \\

\midrule
\multirow{4}{*}{Qwen3-8B}
    & Combined        & $0.1136 \pm 0.0478$          & $\bm{0.5909 \pm 0.0741}$ & $0.0682 \pm 0.0380$ \\
    & PKG    & $\bm{0.4528 \pm 0.0684}$ & $0.5849 \pm 0.0677$          & $\bm{0.3585 \pm 0.0659}$ \\
    & MKG    & $0.2927 \pm 0.0711$          & $0.3415 \pm 0.0741$          & $0.2439 \pm 0.0671$ \\
    & Nothing         & $0.2745 \pm 0.0625$          & $0.3333 \pm 0.0660$          & $0.2157 \pm 0.0576$ \\

\bottomrule
\end{tabular}
\end{table}

In \Cref{tab:supp}, we report the results of an ablation study for evaluating the CenFLARKO models with different input configurations. Notably for the smaller models (Qwen3-0.6B and Qwen3-1.7B), performance decreased when both transaction (PKG) and asset price history (MKG) were combined, relative to using either data source alone. This degradation is likely due to their limited ability to handle increased context lengths with yarn, unlike the larger 4B and 8B models. 

However, even the smaller models still perform better with some data than without (except for Prof@3 and Comb@3 for Qwen3-1.7B), but still not by a wide margin, suggesting that the models have generalized key financial patterns during training and can apply them even in the absence of explicit context.

When comparing the effect of the PKG versus the MKG, the models mostly performed better with the PKG (with exceptions in Qwen3-1.7B and the profitability metric for Qwen3-0.6B). 
The following factors may explain the relative advantage of including the PKG data:
\begin{inparaenum}[(i)]
\item Transaction history is shorter in the PKG and more semantically structured compared to the asset price history in the MKG, allowing the model to more effectively parse and use this input. 
\item LLMs are more adept at capturing human behavioral patterns embedded in transaction logs in the PKG, which resemble their pretraining distribution more closely than numerical price sequences available in the MKG.
\item The aforementioned generalized financial patterns reduce the marginal value of explicit price history in the MKG during inference compared to the highly specific and novel Transactions data specific to each user.
\end{inparaenum}

Interestingly, performance does not scale monotonically with model size. Qwen3-0.6B consistently underperforms, but beyond that, the relationship between scale and effectiveness flattens. Qwen3-1.7B achieves the best scores on Pref@3 and Comb@3, while Qwen3-4B achieves the best Prof@3. The largest model, Qwen3-8B, does not outperform its smaller counterparts on any metric, suggesting that for this domain-specific task, mid-sized models (1.7B–4B) offer the best trade-off between capacity and alignment efficiency. These results underscore that scaling alone is insufficient and that model architecture and fine-tuning strategies may be more critical for real-world financial asset recommendation performance.

\subsection{FedFLARKO Results}

\begin{table*}[t]
\begin{minipage}[b]{0.49\textwidth}
    \centering
    \footnotesize
    \setlength{\tabcolsep}{3pt}
    \caption{Non-IID FedFLARKO Results\\
    \normalfont\small Results are presented as mean $\pm$ standard error of a proportion. Each row corresponds to a Qwen3 model variant trained on a specific subset of user context: combined (PKG + MKG), PKG only (transaction data), MKG only (asset price history), or nothing. The best results for each model are in \textbf{bold}, and the best results overall are marked with a $\uparrow$.}
    \label{tab:supp_fed}
    \begin{tabular}{l@{\hskip 4pt}l@{\hskip 1pt}c@{\hskip 1pt}c@{\hskip 1pt}c}
    \toprule
    \textbf{Model} & \textbf{Data} & \textbf{Pref@3} & \textbf{Prof@3} & \textbf{Comb@3} \\
    \midrule

    \multirow{4}{*}{Qwen3-0.6B}
        & Combined        & $0.0338 \pm 0.0126$         & $0.0628 \pm 0.0169$         & $0.0290 \pm 0.0117$ \\
        & PKG    & $\bm{0.3524 \pm 0.0330}$ & $0.4048 \pm 0.0339$         & $0.2476 \pm 0.0298$ \\
        & MKG    & $0.3381 \pm 0.0326$         & $\bm{0.4619 \pm 0.0344}$ & $\bm{0.2810 \pm 0.0310}$ \\
        & Nothing         & $0.3285 \pm 0.0326$         & $0.4300 \pm 0.0344$         & $0.2705 \pm 0.0309$ \\
        
    \midrule
    \multirow{4}{*}{Qwen3-1.7B}
        & Combined        & $0.0743 \pm 0.0184$         & $\bm{0.4597 \pm 0.0343}$ & $0.0300 \pm 0.0121$ \\
        & PKG    & $\bm{0.3909 \pm 0.0348}$$\uparrow$ & $0.4286 \pm 0.0353$ & $\bm{0.2513 \pm 0.0311}$ \\
        & MKG    & $0.3632 \pm 0.0339$         & $0.4378 \pm 0.0350$         & $0.2289 \pm 0.0296$ \\
        & Nothing         & $0.3283 \pm 0.0334$         & $0.4040 \pm 0.0349$         & $0.2020 \pm 0.0285$ \\

    \midrule
    \multirow{4}{*}{Qwen3-4B}
        & Combined        & $0.3235 \pm 0.0567$         & $\bm{0.6176 \pm 0.0589}$$\uparrow$ & $0.2353 \pm 0.0514$ \\
        & PKG    & $\bm{0.3878 \pm 0.0696}$ & $0.4694 \pm 0.0713$         & $\bm{0.3469 \pm 0.0680}$$\uparrow$ \\
        & MKG    & $0.1915 \pm 0.0574$         & $0.2553 \pm 0.0636$         & $0.1277 \pm 0.0487$ \\
        & Nothing         & $0.2292 \pm 0.0607$         & $0.3958 \pm 0.0706$         & $0.2292 \pm 0.0607$ \\

    \bottomrule
    \end{tabular}
\end{minipage}
\hfill
\begin{minipage}[b]{0.49\textwidth}
    \centering
    \footnotesize
    \setlength{\tabcolsep}{3pt}
    \caption{IID FedFLARKO Results \\
    \normalfont\small 
    Results are presented as mean $\pm$ standard error of a proportion. Each cell is color-coded based on the change in performance relative to the corresponding result in the non-IID setting in \textbf{\Cref{tab:supp_fed}}: \colorbox{red!60!white}{red} indicates a decrease and \colorbox{green!60!white}{green} indicates an improvement, with intensity reflecting the magnitude of the change. 
    Best model results are in \textbf{bold}; best overall are marked with a $\uparrow$.}
    \label{tab:supp_unbiased_fed}
    \begin{tabular}{l@{\hskip 4pt}l@{\hskip 4pt}c@{\hskip 4pt}c@{\hskip 4pt}c}
    \toprule
    \textbf{Model} & \textbf{Data} & \textbf{Pref@3} & \textbf{Prof@3} & \textbf{Comb@3} \\
    \midrule

    \multirow{4}{*}{Qwen3-0.6B}
        & Combined        & \cellcolor{red!17!white}$0.0303 \pm 0.0133$ & \cellcolor{green!19!white}$0.0727 \pm 0.0202$ & \cellcolor{red!20!white}$0.0182 \pm 0.0104$ \\
        & PKG    & \cellcolor{green!26!white}$0.3750 \pm 0.0365$ & \cellcolor{green!37!white}$0.4545 \pm 0.0375$ & \cellcolor{red!21!white}$0.2330 \pm 0.0319$ \\
        & MKG    & \cellcolor{green!45!white}$\bm{0.4111 \pm 0.0367}$$\uparrow$ & \cellcolor{red!26!white}$0.4333 \pm 0.0369$ & \cellcolor{red!27!white}$0.2611 \pm 0.0327$ \\
        & Nothing         & \cellcolor{green!41!white}$0.3880 \pm 0.0360$ & \cellcolor{green!45!white}$\bm{0.5027 \pm 0.0370}$$\uparrow$ & \cellcolor{green!22!white}$\bm{0.2896 \pm 0.0335}$ \\
        
    \midrule
    \multirow{4}{*}{Qwen3-1.7B}
        & Combined        & \cellcolor{red!21!white}$0.0597 \pm 0.0167$ & \cellcolor{red!62!white}$0.3524 \pm 0.0330$ & \cellcolor{red!17!white}$0.0251 \pm 0.0111$ \\
        & PKG    & \cellcolor{red!22!white}$0.3750 \pm 0.0415$ & \cellcolor{red!19!white}$0.4203 \pm 0.0420$ & \cellcolor{green!21!white}$0.2647 \pm 0.0378$ \\
        & MKG    & \cellcolor{green!33!white}$\bm{0.4015 \pm 0.0419}$ & \cellcolor{green!20!white}$\bm{0.4493 \pm 0.0423}$ & \cellcolor{green!40!white}$\bm{0.2847 \pm 0.0386}$ \\
        & Nothing         & \cellcolor{green!22!white}$0.3448 \pm 0.0395$ & \cellcolor{green!16!white}$0.4069 \pm 0.0408$ & \cellcolor{green!47!white}$0.2759 \pm 0.0371$ \\

    \midrule
    \multirow{4}{*}{Qwen3-4B}
        & Combined        & \cellcolor{red!45!white}$0.2532 \pm 0.0489$ & \cellcolor{red!36!white}$\bm{0.5696 \pm 0.0557}$ & \cellcolor{red!52!white}$0.1519 \pm 0.0404$ \\
        & PKG    & \cellcolor{red!29!white}$\bm{0.3559 \pm 0.0623}$ & \cellcolor{red!28!white}$0.4407 \pm 0.0646$ & \cellcolor{red!33!white}$\bm{0.3051 \pm 0.0599}$$\uparrow$ \\
        & MKG    & \cellcolor{green!33!white}$0.2308 \pm 0.0584$ & \cellcolor{green!55!white}$0.3462 \pm 0.0660$ & \cellcolor{green!26!white}$0.1538 \pm 0.0500$ \\
        & Nothing         & \cellcolor{red!78!white}$0.0833 \pm 0.0399$ & \cellcolor{red!100!white}$0.1667 \pm 0.0538$ & \cellcolor{red!88!white}$0.0625 \pm 0.0349$ \\

    \bottomrule
    \end{tabular}
\end{minipage}
\end{table*}

In \Cref{tab:supp_fed}, we demonstrate the results of our federated FLARKO models fine-tuned across the clients with non-IID data. Our results here are overall similar to the centralized results, but with a decrease in performance. However, we see that Qwen3-4B's relative performance is better here as it achieves the highest Comb@3 score, on top of the highest Prof@3 score. As can be easily seen in \Cref{fig:baselines}, Qwen3-4B manages to improve its results in the federated testing compared to its centralized counterpart, specifically in its peak Pref@3 and Comb@3 scores. This indicates that the larger models may be more resilient to the performance degradation caused by federation or may even benefit from it.

In \Cref{tab:supp_unbiased_fed}, we show the federated model results when fine-tuned across clients with IID data. The results highlighted in green are better than the corresponding non-IID client scores, and the reverse is true for the scores highlighted in red. We observe that the two smaller models mostly have better results with IID clients, notably with their peak scores (except for Qwen3-1.7B's Prof@3 peak score being slightly lower). However, Qwen3-4B surprisingly performs much better with non-IID clients. Notably, all three of its peak scores are higher with the non-IID clients. 
This indicates that Qwen3-4B is not only resistant to non-IID data but instead thrives with it. This further supports FLARKO's performance in a realistic, federated setting.

\subsection{Baselines Comparison}

In \Cref{fig:baselines} from earlier in the paper, we show our best results from \Cref{tab:supp} and \Cref{tab:supp_fed} against the results of other models introduced in \Cref{subsec:baselines}. 

Random Forest, Linear Regression, and LightGBM use the asset price history (non-behavioral MKG data), while Popularity, LightGCN, ARM, MF, and UB kNN use the customer transaction history (behavioral PKG data). The random baseline uses neither. While most of the baseline models (except Random, Random Forest, and Popularity) outperformed our models in Prof@3, our models outperformed them in both Pref@3 and Comb@3. 

Random Forest, Linear Regression, LightGBM excel in pure profitability, due to their direct focus on financial returns without the added complexity of derived human preferences from the transaction data. However, profitable recommendations may not necessarily be good recommendations, as users will ignore recommendations that do not align with their preferences. On the other hand, even our weakest models significantly outperform all the baseline models at Comb@3 by a wide margin. This is the most important metric as it specifically looks for recommendations that are profitable and align with the user's preferences. Overall, we show that both CenFLARKO and FedFLARKO perform significantly better than their competition.

\section{Conclusion}

We introduce FLARKO, a unified framework for financial asset recommendation that delivers behaviorally aligned outputs by grounding LLM reasoning in structured KGs. FLARKO supports both centralized (CenFLARKO) and federated (FedFLARKO) deployments, offering scalable fine-tuning in low-resource settings and collaboration across institutions.
This work lays the foundation for a new generation of financial asset recommendation systems that are intelligent and fundamentally aligned with the users and institutions they serve, meeting the growing demand for personalization without compromising trust or compliance.

\begin{sloppypar}
    Through comprehensive experiments, we demonstrate that FLARKO consistently outperforms traditional and LLM-based baselines by generating recommendations that are not only profitable but also behaviorally aligned.
Ablation studies further validate FLARKO's design: both historical market context (MKG) and personalized behavioral signals (PKG) contribute meaningfully to recommendation quality, with their combination yielding the strongest performance.
\end{sloppypar}

The success of our FL approach under non-IID client distributions demonstrates a scalable path toward collaborative AI systems. Moreover, our exploration of PKG adaptation reveals a powerful mechanism for encoding behavioral constraints and portfolio-level objectives directly into the model’s symbolic reasoning process. This enables FLARKO to move beyond static personalization, actively steering recommendations toward diversified, goal-aligned investment strategies.

Looking ahead, we aim to enhance these capabilities by incorporating richer behavioral signals, more expressive constraint templates, and real-time user feedback, advancing toward a dynamic financial assistant that is both adaptive and aligned with investor goals.

\section{Online Resources}
\label{sec:resources}

All research artifacts, including source code, dataset construction scripts, and result generation pipelines, are available in our GitHub repository. All external datasets and software dependencies used in this work are documented and linked in the repository’s README.\\
\url{https://github.com/brains-group/FLARKO}.



\balance
\bibliographystyle{ACM-Reference-Format}
\bibliography{references}

\end{document}